# Tuning band alignment at a semiconductor-crystalline oxide heterojunction via electrostatic modulation of the interfacial dipole

M. Chrysler[1], J. Gabel[2], T.-L. Lee[2], A. N. Penn[3], B. E. Matthews[4], D. M. Kepaptsoglou[5,6], Q. M. Ramasse[5,7], J. R. Paudel[8], R. K. Sah[8], J. D. Grassi[8], Z. Zhu[9], A. X. Gray[8], J. M. LeBeau[10], S. R. Spurgeon[4], S. A. Chambers[11], P. V. Sushko[11], and J. H. Ngai[1]

[1]Department of Physics, University of Texas-Arlington, Arlington, TX 76019, USA
[2] Diamond Light Source, Ltd., Harwell Science and Innovation Campus, Didcot OX11 0DE, UK
[3]Department of Materials Science & Engineering, North Carolina State University, Raleigh, NC 27695, USA
[4]Energy and Environment Directorate, Pacific Northwest National Laboratory, Richland, WA 99352, USA
[5]SuperSTEM, Daresbury, Warrington WA4 4AD, UK
[6]Department of Physics, University of York, York YO10 5DD, UK
[7]School of Physics and Astronomy & School of Chemical and Process Engineering, University of Leeds, Leeds LS2 9JT, UK
[8]Department of Physics, Temple University, Philadelphia, 19122 PA, USA
[9] Environmental Molecular Sciences Laboratory, Earth and Biological Sciences Directorate, Pacific Northwest National Laboratory, Richland, WA 99352, USA
[10]Department of Materials Science & Engineering, Massachusetts Institute of Technology, Cambridge, MA 02139, USA
[11] Physical Sciences Division, Physical & Computational Sciences Directorate, Pacific Northwest National Laboratory, Richland, WA 99352, USA

We demonstrate that the interfacial dipole associated with bonding across the $SrTiO_3$/Si heterojunction can be tuned through space charge, thereby enabling the band alignment to be altered via doping. Oxygen impurities in Si act as donors that create space charge by transferring electrons across the interface into $SrTiO_3$. The space charge induces an electric field that modifies the interfacial dipole, thereby tuning the band alignment from type-II to type-III. The transferred charge, resulting in built-in electric fields, and change in band alignment are manifested in electrical transport and hard x-ray photoelectron spectroscopy measurements. Ab initio models reveal the interplay between polarization and band offsets. We find that band offsets can be tuned by modulating the density of space charge across the interface. Functionalizing the interface dipole to enable electrostatic altering of band alignment opens new pathways to realize novel behavior in semiconducting heterojunctions.

# I. INTRODUCTION

Semiconducting heterojunctions that exhibit built-in electric fields stemming from a net transfer of itinerant charge are the building blocks for virtually all electronic device technologies. Recent advancements have enabled charge transfer and built-in fields to be studied across atomically abrupt and structurally coherent interfaces between semiconductors and crystalline complex oxides [1, 2]. Such heterojunctions provide a fundamentally novel setting to study charge transfer, as they exhibit discontinuities in composition and crystal structure across the interface, and the bonding transitions from being covalent to ionic in nature. In terms of applications, the mixed covalent and ionic characteristics of hybrid heterojunctions could enable functionality that cannot be achieved using either material alone [3-5].

Band alignment plays a key role in determining the direction of charge transfer across a heterojunction. In general, band offsets are determined not only by differences in bulk properties (e.g., electron affinities), but also the distribution of charge across the interface due to bonding. In this regard, a hallmark feature of a crystalline oxide that is epitaxially grown on a covalent semiconductor is the formation of a robust dipole at the interface that is induced by the strong electronegativity of O. For the archetypal heterojunction between $SrTiO_3$ (STO) and Si, O anions in the STO that are adjacent to the interface create a dipole by pulling electrons away from Si atoms at the surface [6-9]. This dipole induces the structure of the STO near the interface to become polarized, as oxygen anions (Sr/Ti cations) are attracted towards (repelled from) the positive interfacial layer of Si.

McKee et al. first recognized the significance of the interfacial dipole in electrostatically altering band offsets at semiconductor-crystalline oxide interfaces [10]. They tuned the band offset between a rocksalt oxide and Si by modifying the dipole via the chemical composition at the interface. While changes in composition represent an *intrinsic* approach to modify the dipole, the electrostatic nature of the system suggests the dipole could also be altered *extrinsically* via electric fields [11]. Introducing dopants is one approach to create local electric fields. Fields arise when itinerant electrons/holes become spatially separated from their immobile donor/acceptor ion cores to form space charge. Electric fields due to space charge could be very large in heterojunctions between STO and Si, as the ability of Ti to exhibit multiple oxidation states gives rise to very short screening lengths.

Here, we demonstrate that the interfacial dipole across the heterojunction between STO and Si can be tuned through space charge, thereby enabling band offsets and ultimately band-alignment (i.e., type-II to type-III) to be modified. We present a comparative study of STO grown on Czochralski (STO/CZ-Si) and float-zone (STO/FZ-Si) Si wafers, which differ in density of O impurities in the Si, which in turn act as n-type donors. The electrical transport of STO/CZ-Si differs from STO/FZ-Si heterojunctions by exhibiting a reduced sheet resistance along with the formation of a hole-gas in the Si at higher temperatures. Hard x-ray photoelectron spectroscopy (HAXPES) measurements reveal built-in fields and a type-III band alignment in the STO/CZ-Si heterojunction, in which the conduction band of STO is below the valence band of Si. In contrast, a type-II alignment and no built-in fields are found in the STO/FZ-Si heterojunction, in which the conduction band of STO is above (below) the valence (conduction) band of Si. The difference in band alignment is correlated with the larger (smaller) concentration of O impurity donors that accumulate below the surface of the CZ-Si (FZ-Si), as revealed by time-of-flight secondary ion mass spectroscopy (ToF-SIMS). These impurity donors pass charge to the STO, creating an electric field across the interface that modifies the dipole and therefore the offset between the valence bands of STO and Si. Ab initio calculations reveal the relationship between the interfacial dipole and band alignment. The band offsets can be tuned by modulating the density of space charge in the Si wafer via the density of n-type O donors. The ability to alter band alignment through space charge sets hybrid heterojunctions apart from conventional heterojunctions and provides a new degree of freedom to realize functional behavior.

## II.     EXPERIMENTAL DETAILS

Epitaxial STO films were grown on (100)-oriented, 2" diameter 300 μm thick Si wafers (Virginia Semiconductor) using reactive molecular beam epitaxy (MBE) in a custom-built chamber operating at a base pressure of < 3 × 10$^{-10}$ Torr. Growth was performed on nominally undoped CZ-Si and FZ-Si wafers, as well as n-type CZ-Si wafers that were doped with phosphorus (MTI). All STO films were grown under identical conditions, as described below. The wafers were introduced into the MBE chamber and cleaned by exposing to activated oxygen generated by a radio frequency source (VEECO) operated at ~ 250 W to remove residual organics from the surface. Two monolayers of Sr were deposited at a substrate temperature of 550 °C, which was subsequently heated to 870 °C to remove the native layer of $SiO_x$ through the formation and desorption of SrO [12]. A 2 × 1 reconstruction was observed in the reflection high energy electron diffraction (RHEED) pattern, indicating a reconstructed Si surface. Next a half monolayer of Sr was deposited at 660 °C to form a template for subsequent layers of STO. The substrate was then cooled to room temperature, at which temperature 2.5 ML of SrO and 2 ML of $TiO_2$ were co-deposited and then heated to 500 °C to form 2.5 unit cells (u.c.) of crystalline $SrTiO_3$. Subsequent layers of STO were grown at a substrate temperature of 580 °C through co-deposition of Sr, and Ti fluxes in a background oxygen pressure of $4 \times 10^{-7}$ Torr. Thermal effusion cells (Veeco and SVT Associates) were used to evaporate Sr and Ti source materials (Alfa Aesar). All fluxes were calibrated using a quartz crystal microbalance (Inficon). Typical growth rates were ~ 1 u.c. per minute.

HAXPES measurements were performed at Beamline I09 at Diamond Light Source (UK). The X-ray energy was tuned to 5.9, 4.05 and 2.4 keV using a Si(111) double crystal monochromator to investigate the 12, 8 and 4 nm STO films, respectively. For the data recorded at 5.9 and 4.05 keV the energy resolution was further improved by a Si(004) and a Si(022) channel-cut high resolution monochromator, respectively. Photoelectrons were collected by a Scienta Omicron EW4000 high-energy hemispherical analyzer with a pass energy of 200 eV, resulting in an overall experimental resolution of ~ 0.25 eV at the three photon energies determined by fitting the Au Fermi edge with a Fermi function. The binding energy scale was calibrated using the Au 4f core levels, along with the Au Fermi edge. The x-ray angle of incidence was about 10° off the sample surface, and the photoelectrons were collected over a range of ±28° centered around the X-ray polarization direction. Additional HAXPES measurements at the photon energy of 5410 eV were carried out with a laboratory-based spectrometer at Temple University using a focused monochromated Cr Kα X-ray source and a similar EW4000 analyzer.

Electrical transport measurements of the heterojunctions were performed in the van der Pauw geometry in a Quantum Design Dynacool$^{TM}$ System as a function of temperature and applied magnetic field. Electrical contacts were established on the four corners of diced 4 mm × 4 mm or 5 mm × 5 mm samples using Al wedge bonding (Westbond). The contacts exhibited linear characteristics in 2-point current-voltage measurements, confirming ohmic behavior. Resistivity and Hall measurements were performed using a Keithley 2400 Sourcemeter in conjunction with a Keithley 2700 fitted with a 7709 matrix module multiplexer that enabled all the van der Pauw lead configurations to be probed. The Hall data was fitted to a two-carrier model that accounts for electron conductivity in the STO in conjunction with either hole or electron conductivity in the Si. Details on the fitting procedure can be found elsewhere [2].

Cross-sectional scanning transmission electron microscopy (STEM) samples of the 12 nm STO/CZ-Si and 12 nm STO/FZ-Si heterojunctions were prepared using FEI Helios NanoLab DualBeam $Ga^+$ Focused Ion Beam (FIB) with a standard lift out procedure, followed by $Ar^+$ ion milling. Monochromated electron energy loss spectroscopy (EELS) measurements were performed at 60kV using the monochromated Nion UltraSTEM 100MC - Hermes instrument, equipped with a Gatan Enfinium ERS spectrometer. The energy resolution was set to ~0.150eV by the monochromator slit for the Ti $L_{2,3}$ edge mapping and a 0.3 eV native width was used for the other maps. The beam convergence angle was 31.5 mrad, and the EELS collection angle was set to 44mrad. Data were acquired in a DualEELS mode and subsequently corrected for energy drift. Composition maps were subsequently filtered using principal component analysis (PCA) to improve signal-to-noise, but spectra for fine structure analysis were unfiltered.

For the 8 nm thick STO/CZ-Si sample, cross-sectional STEM samples were prepared by conventional wedge polishing and $Ar^+$ ion milling with a Fischione 1050 TEM Mill cooled with liquid nitrogen. STEM imaging was performed on a probe corrected Thermo Fisher Titan G2 operated at 200 kV. Annular Dark Field (ADF) and integrated differential phase contrast (iDPC) images were acquired with the Revolving STEM (RevSTEM) method to maximize signal to noise and minimize the effects of sample drift and scan distortion [13]. Atom column positions were measured using Atom Column Indexing and custom Python codes [14]. The polarization was measured as the displacement of oxygen atom columns relative to the center position of the u.c., shown in the inset of Fig. 1(c).

ToF-SIMS measurement were performed with a TOF.SIMS5 instrument (IONTOF GmbH, Münster, Germany) using a dual beam depth profiling strategy. A 1.0 keV $Cs^+$ beam (~ 45 nA) was used for sputtering. The $Cs^+$ beam was scanned over a 300 × 300 $\mu m^2$ area. A 25.0 keV $Bi_3^+$ beam (~ 0.57 pA) was used as the analysis beam to collect SIMS depth profiling data. The $Bi_3^+$ beam was focused to be ~ 5 μm in diameter and scanned over a 100 × 100 $\mu m^2$ area at the center of the $Cs^+$ crater.

For the *ab initio* simulations based on the density functional theory (DFT), the STO/Si heterojunction was represented using the periodic slab model. The Si part of the slab was 19 atomic planes thick; it was terminated with (001) planes with the lateral cell corresponding to the $\sqrt{2}\times\sqrt{2}$ bulk diamond crystallographic cell and 2×2 bulk perovskite cell. Dangling bonds on the one side of the Si slab were saturated with hydrogen atoms. The thickness of the simulated STO film on the other side of the slab was varied between 1 and 9 u.c., terminated with $TiO_2$ plane in each case. Here we adopted the interface structure reported earlier in Ref [8] that is consistent with the result of the STEM analysis of the samples used in this study. The lateral cell parameters of the slab supercell were fixed at the values a = b = 7.68 Å, while the off-interface parameter was fixed at 90 Å; this leaves a vacuum gap of ~ 29 Å or larger depending on the STO film thickness. The total energy of each system was minimized with respect to the internal coordinates. The calculations were performed using the VASP package [15, 16] and the PBEsol density functional [17]. The projector-augmented wave potentials were used to approximate the effect of the core electrons [18]. A Γ-centered 2×2×1 *k*-mesh was used for Brillouin-zone integration in the structure optimization calculations; a 12×12×1 *k*-mesh was used for calculations of the density of states (DOS). The plane-wave basis-set cutoff was set to 500 eV. The total energy convergence criterion was set to $10^{-5}$ eV. The charge population analysis was performed using the method developed by Bader [19, 20]. One-electron DOS was smeared by convoluting it with Gaussian functions with the full width at half maximum of 0.2 eV.

I.   **RESULTS & DISCUSSION**

STEM imaging reveals the structural polarization of the STO near the interface that accompanies the formation of the interfacial dipole. Figures 1(a) and 1(b) show ADF and iDPC images of the 8 nm STO/CZ-Si heterojunction, respectively. Polarization manifested as downward displacements of O anions relative to Sr and Ti cations can be seen in the iDPC image of Fig. 1(c). The magnitude of these displacements, which are mapped in Fig. 1(d) and Fig. S1, are most pronounced near the interface, as summarized in Fig. 1(e). The rapid decrease in polarization away from the interface can be attributed to structural relaxation of the STO relative to the Si substrate. This structural relaxation is driven by anti-phase boundaries (Fig. S2) that nucleate at steps on the Si surface at which the step-heights are incommensurate with the out-of-plane lattice constant of STO [21].

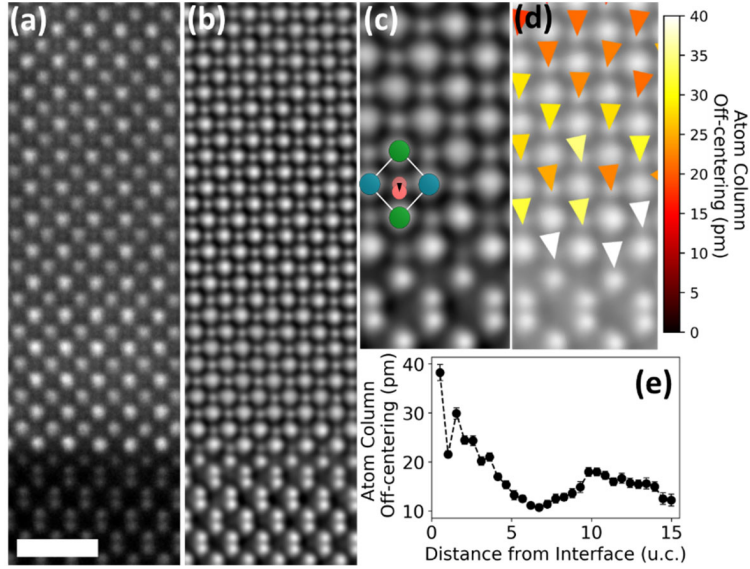

FIG. 1. (a) ADF image of the 8 nm STO/CZ-Si heterojunction. Scale bar is 1 nm (b) iDPC image of the 8 nm STO/CZ-Si heterojunction. (c) Expanded view of the iDPC image. Polarization of the STO near the interface is manifested as O anions (pink) are displaced downwards relative to Sr (cyan) and Ti (green) cations. (d) Corresponding map of oxygen column displacement. (e) Average displacement of oxygen anions at each atomic layer.

To better understand the character of the interfaces, we have also performed atomically-resolved EELS measurements on the 12 nm STO/CZ-Si and 12 nm STO/FZ-Si FZ heterojunctions. As shown in Fig. 2(a) and 2(b), mapping of the Sr $L_{2,3}$, and Si $K$ edges reveals a sharp interface; these edges were specifically chosen for the best spatial localization, revealing an abrupt chemical profile for both the Sr and Si signals. Focusing specifically on the Sr signal, we note that the terminal layer of the STO appears to be an SrO plane in both heterojunctions. Figures 2(c) and (d) show a second map with the spectrometer optimized to measure the fine structure of the Ti $L_{2,3}$ edge of the 12 nm STO/CZ-Si and 12 nm STO/FZ-Si heterojunctions, respectively. In particular, we compare changes in the $t_{2g}$ / $e_g$ splitting of the $L_3$ and $L_2$ edges, which is a known indicator of local valence state [22]. In the bulk region (spectra 6), we observe a clear separation among all four features for both the 12 nm STO/CZ-Si and 12 nm STO/FZ-Si heterojunctions. In the case of the 12 nm STO/CZ-Si, we measure a slight decrease in the separation and intensity ratio of the $L_3$ edge features beginning at the fourth $TiO_2$ plane of the

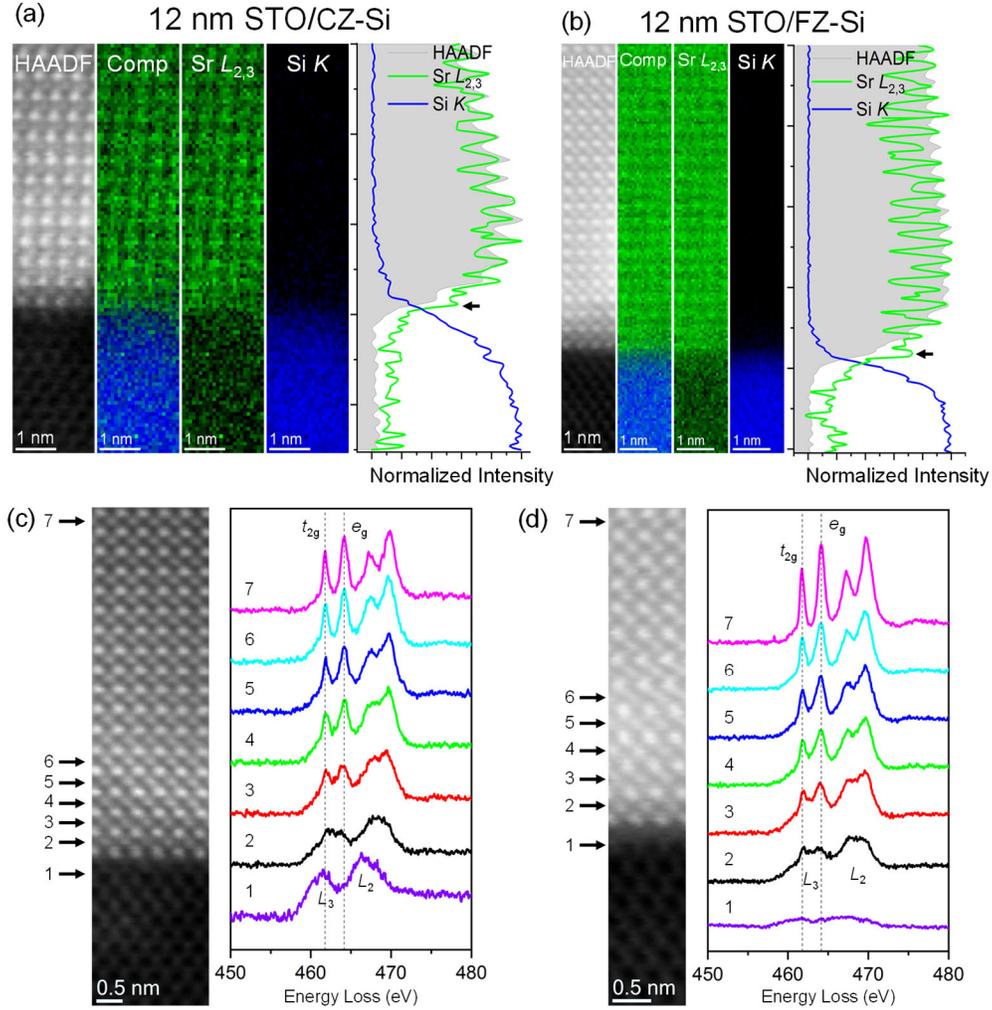

FIG. 2. (a) and (b) STEM-EELS mapping of the Sr $L_{2,3}$, and Si $K$ edges for the 12 nm STO/CZ-Si and 12 nm STO/FZ-Si heterojunctions, respectively. (c) and (d) STEM-EELS fine structure of the Ti $L_{2,3}$ edge of the 12 nm STO/CZ-Si and 12 nm STO/FZ-Si heterojunctions, respectively.

STO layer. The separation continues to decrease plane-by-plane until the $t_{2g}$ and $e_g$ nearly merge at the TiO$_2$ plane closest to the Si. This trend points to slight Ti$^{3+}$-like character in the 3–4 TiO$_2$ planes closest to the interface, relative to the bulk STO spectrum. On the Si side of the interface there is a significant Ti$^{0+}$-like signal, consistent with either some degree of interdiffusion, or steps on the Si surface within the cross-section of the sample. In contrast, for the 12 nm STO/FZ-Si sample, the bulk line shape is preserved up to the 1–2 planes closest to the interface, where slight merging occurs, again consistent with some Ti$^{3+}$-like character. A nearly negligible Ti signal is present on the Si of the interface in this case, as is shown in the multiple linear least squares (MLLS) fitting in Fig. S3. This difference in Ti valence between the heterojunctions is further probed using HAXPES, as discussed below.

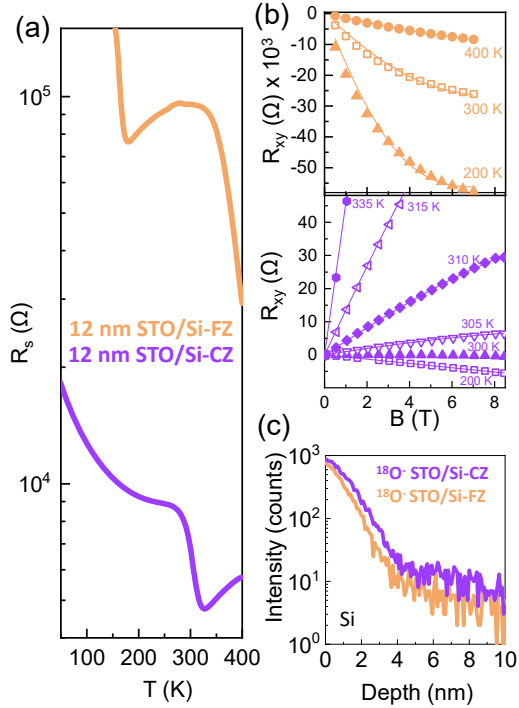

FIG. 3. (a) $R_s$ for the 12 nm STO/Si-FZ and 12 nm STO/CZ-Si heterojunctions. (b) $R_{xy}$ for the 12 nm STO/Si-FZ (orange) and 12 nm STO/CZ-Si (purple) heterojunctions. Note the crossover in sign of $R_{xy}$ in the latter indicating the formation of a hole-gas. Raw data are shown as symbols, whereas fits to the data using a 2-carrier model are shown as lines. (c) $^{18}O^-$ ToF-SIMS depth profiles in Si for the 12 nm STO/Si-CZ and 12 nm STO/FZ-Si heterojunctions showing higher density of $^{18}O^-$ impurities in the former.

Despite identical growth conditions and interfacial structures, however, the 12 nm STO/FZ-Si and 12 nm STO/CZ-Si heterojunctions exhibit markedly different electrical transport characteristics. The 12 nm STO/FZ-Si heterojunction exhibits non-monotonic behavior in the sheet resistance ($R_s$) during cooling. $R_s$ increases from 400 K to 300, gradually decreases from 300 K to 179 K, and then rapidly increases again for $T < 179$ K, as shown in Fig. 3(a). The Hall resistance $R_{xy}$ in the STO/FZ-Si heterojunction is negative and non-linear throughout the entire temperature regime of measurement (Fig. 3(b)). The non-linearity arises from electrons in the Si exhibiting higher mobilities, as discussed below. In comparison, the STO/CZ-Si heterojunction exhibits a lower $R_s$ throughout the entire temperature range of measurement, as well as a much more pronounced non-monotonic anomaly near ~ 300 K (Fig. 3(a)). The anomaly arises from the formation of a robust hole-gas in the Si that conducts in parallel with electrons in the STO above ~ 300 K, as indicated by non-linearity and a crossover in sign of $R_{xy}$ from negative to positive with increasing temperature (Fig. 3(b)]. The lower $R_s$ in the 12 nm STO/CZ-Si heterojunction relative to the 12 nm STO/FZ-Si heterojunction stems from a higher electron sheet density $n_{e(STO)}$ in the STO. Hall measurements at ~ 200 K, below the temperature at which a hole-gas forms, indicates $n_{e(STO)}$ ~ $9.7 \times 10^{14}$ cm$^{-2}$ in the STO of the STO/CZ-Si heterojunction. In comparison, $n_{e(STO)}$ for the STO/FZ-Si heterojunction at ~ 200 K is more than an order of magnitude lower, ~ $3.3 \times 10^{13}$ cm$^{-2}$. Residual oxygen vacancies cannot explain the larger $n_{e(STO)}$ measured in the STO of the STO/CZ-Si heterojunction, as both STO/CZ-Si and STO/FZ-Si heterojunctions were grown under identical conditions. In this regard, the sheet density of carriers in the STO of the 12 nm STO/FZ-Si heterojunction represents an *upper bound* to the number of carriers generated due to residual oxygen vacancies in either heterojunction.

HAXPES measurements at 300 K for the Ti 2p core-level spectrum corroborate the larger $n_{e(STO)}$ in the 12 nm STO/CZ-Si heterojunction relative to the 12 nm STO/FZ-Si heterojunction. Figure 4(a) shows the Ti 2p spectra obtained for the 12 nm STO/CZ-Si and 12 nm STO/FZ-Si heterojunctions, along with the spectrum measured on a bulk single-crystal of Nb-doped SrTiO$_3$ for comparison. Note the pronounced intensities of the lower valence features of Ti (i.e., Ti$^{3+}$, Ti$^{0+}$ etc.) in the spectrum from the STO/CZ-Si heterojunction, consistent with the large $n_{e(STO)}$ that is measured. Angle-resolved HAXPES and STEM-EELS measurements indicate the lower valence features reside near the interface [2]. In contrast, the lower valence features in the core-level spectra of Ti for the 12 nm STO/FZ-Si heterojunction exhibit much weaker intensity, consistent with the lower $n_{e(STO)}$ measured. These results are corroborated by lab-based HAXPES measurements performed at Temple University (see Fig. S4).

Equally intriguing, the redistribution of intensity in all core-level spectra for the 12 nm STO/CZ-Si heterojunction (see arrows in Fig. 4(a)] relative to spectra for the phase pure component materials is consistent with the presence of built-in electric fields. The potential drops associated with these electric fields result in a distribution of binding energies in the affected layers, thereby perturbing the volume-averaged line shapes that are measured. To extract the built-in potentials, we fit the Si 2p, Ti 2p$_{3/2}$ and Sr 3d spectra for the 12 nm STO/CZ-Si heterojunctions to sums of reference spectra. Each reference spectrum is taken from a pure, bulk crystal of Si or STO that is minimally affected by surface core-level shifts and band bending, and is assigned to the different layers in the heterojunction. As described in detail elsewhere [2, 23], the spectral weights are scaled to account for the depths of the layers to which they are assigned, and their binding energies are systematically varied to explore all possible values of the built-in potential, subject to the constraint of monotonic variation with depth. This fitting procedure yields a depth-resolved map of the band bending and band offsets across the heterojunction.

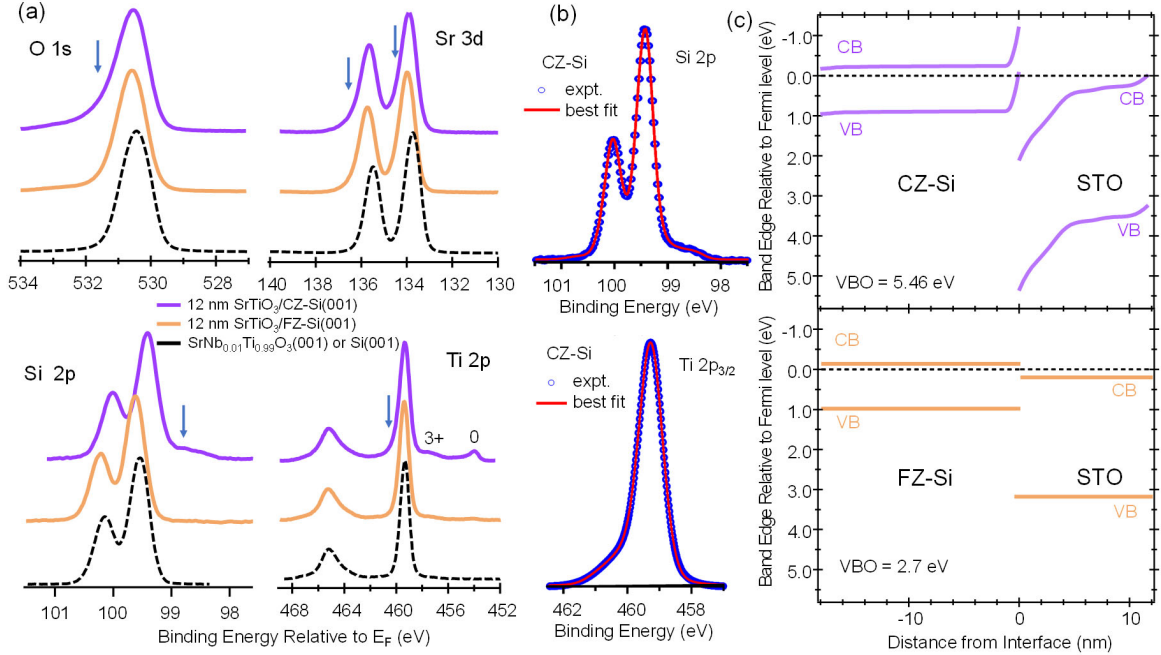

FIG. 4. (a) Core-level O 1s, Sr 3d, Si 2p, and Ti 2p, spectra for the 12 nm thick STO/CZ-Si (purple) and STO/FZ-Si (orange) heterojunctions. Reference spectra from Si(001) and SrNb$_{0.01}$Ti$_{0.99}$O$_3$(001) single crystal substrates are also shown. Arrows denote asymmetries observed in the 12 nm STO/CZ-Si heterojunction. (b) Fits to the asymmetric Si 2p and Ti 2p$_{3/2}$ spectra of the 12 nm STO/CZ-Si heterojunction. (c) Band-offsets across the 12 nm STO/CZ-Si and STO/FZ-Si heterojunctions. Note the type-III (type-II) alignment in the STO/CZ-Si (STO/FZ-Si) heterojunction.

In the upper panel of Fig. 4(c) we show the valence ($E_V$) and conduction band ($E_C$) edge energies as a function of distance from the interface, as extracted from the fits shown in Fig. 4(b). For both Si and STO in the STO/CZ-Si heterojunction, the valence band edge relative to the Fermi level is given by $E_V(z) = E_{CL}(z) - (E_{CL} - E_V)_{ref}$ in which $E_{CL}(z)$ is the core-level binding energy vs. depth $z$ derived from the fits, and $(E_{CL} - E_V)_{ref}$ is the difference in energy between the same core level and the valence band maximum for the appropriate bulk single-crystal reference sample. The conduction band (CB) edge, $E_C(z)$, is given by $E_V(z) - E_g$, where $E_g$ is the bulk bandgap. Once the band-edge profiles are extracted from the core-level fits (Fig. 4(c)), the band offsets are determined simply by taking differences of band-edge energies directly at the interface. Thus, the valence band offset (VBO) can be expressed as $\Delta E_V = E_V^{STO}(z=0) - E_V^{Si}(z=0)$ and $\Delta E_C = E_g^{STO} - E_g^{Si} - \Delta E_V$. A value of 5.46(16) eV results for the band-edge profiles shown for 12 nm STO/CZ-Si in Fig. 4(c), indicating the presence of a type-III band alignment. In contrast, the core-level spectra of the STO/FZ-Si heterojunction exhibits much less asymmetry or broadening. The STO/FZ-Si exhibits a type-II alignment in which the valence band-offset is 2.7(1) eV.

Despite the type-III band alignment of the STO/CZ-Si heterojunction, the valence band of Si is not the principal source of itinerant electrons in the STO. Fits to the Hall data using a two-carrier model indicate that the sheet density of holes in the Si is much lower than that for electrons in the STO (i.e. $n_{h(Si)} \ll n_{e(STO)}$, see Fig. S5). The conductivities of the hole and electron gases are comparable since the mobility of holes in the Si far exceeds that of electrons in the STO (i.e. $\mu_{h(Si)} \gg \mu_{e(STO)}$). Significantly, ToF-SIMS reveals the presence of a high density of O impurities in the Si near the interface (Figure 3(c)). This observation is also consistent with the non-negligible concentration of O observed in Si via STEM-EELS (Figs. S7, S8). Although the Si in STO/CZ-Si is nominally undoped, CZ grown Si naturally contains O impurities because it is crystallized in a silica crucible. The O diffuses toward the Si surfaces at elevated temperatures and becomes an n-type donor [24]. As discussed below, O also diffuses into the Si substrate during epitaxial growth. In the presence of a type-III band alignment, electrons from O donors in Si readily transfer to the STO, forming a depletion region of positive space charge in the former, and an electron gas in the latter. Thus, the hole-gas in the Si is a manifestation of inversion. Consistent with this picture, the built-in potential generated by the spatial profile of ionized O impurities measured by ToF-SIMS matches the spatial profile of the upward band-bending determined from HAXPES exceedingly well (see Ref. [2]).

ToF-SIMS reveals that O impurities also exist near the interface in the Si of the STO/FZ-Si heterojunction, albeit at lower densities (Figure 3(c)). As O is not intrinsically dissolved into the bulk of FZ grown Si crystals, we infer that these O impurities diffuse into the Si during epitaxial growth. The higher densities of O donors in the Si of the STO/CZ-Si heterojunction is attributed to the additional contribution from native O impurities. The O impurities in the STO/FZ-Si heterojunction also act as n-type donors, like those in the STO/CZ-Si heterojunction. However, in contrast to the STO/CZ-Si heterojunction, the electrons associated with the O-donors in the Si of the STO/FZ-Si heterojunction are not fully depleted through transfer to STO, but rather partially contribute to n-type conductivity in the FZ-Si. Hence, the non-linearity in $R_{xy}$ for the 12 nm STO/FZ-Si heterojunction (Fig. 3(b)] stems from these itinerant electrons that remain in the FZ-Si, as they exhibit much higher mobilities than the electrons in the STO. Fits to $R_{xy}$ of the STO/FZ-Si heterojunction indicate that the sheet density of electrons in the FZ-Si is $n_{e(Si)} \sim 1.2 \times 10^{10}$ cm$^{-2}$, at 300 K (Fig. S5).

We now consider how a higher density of O donors can give rise to a type-III band-alignment that drives depletion and inversion in the 12 nm STO/CZ-Si heterojunction. We propose that electron transfer becomes *self-reinforcing* in the presence of the interfacial dipole. As itinerant electrons are transferred from Si to STO, the electric field associated with space charge modifies the interfacial dipole such that the offset between the valence bands of STO and Si increases, which in turn promotes additional electron transfer. A transition in band alignment from type-II to type-III is induced if a sufficient density of space charge accumulates due to electron transfer.

First principles DFT calculations reveal insights to this picture. We first elucidate the interplay between polarization in the STO and the valence band offset by comparing polar and non-polar STO/Si heterojunctions in which the Si is initially undoped. For the non-polar scenario, atoms of the SrO and TiO$_2$ planes within the STO/Si heterojunction are artificially fixed at the positions corresponding to their average respective positions in fully relaxed bulk STO. For the polar scenario, we define polarization as the difference between average positions of cations (C) and anions (A) in the off-plane direction ($z$) in each atomic plane (see Fig.5(a)]. Within the first SrO plane at the interface, the out-of-plane separation between the Sr$^{2+}$ and O$^{2-}$ ions is driven by competition between the formation of polar Si-O$^{2-}$ bonds, and attraction between Sr$^{2+}$ and O$^{2-}$ ions. While interfacial Si-O$^{2-}$ bonds form for both polar and non-polar heterojunctions, the positions of atoms in the former are allowed to relax. Consequently, cations (anions) in the STO are repelled from (attracted to) the positive interfacial Si, giving rise to polarization in the +$z$ direction. However, surface rumpling, whereby larger O$^{2-}$ ions displace outward and smaller Ti$^{4+}$ ions displace inward, favors polarization of the film in the -z direction. The competition between the interface and surface in respectively inducing +z and -z oriented polarizations, results in a polarization that spatially varies throughout the film, as shown in Fig.5(b) for various film thicknesses $N$.

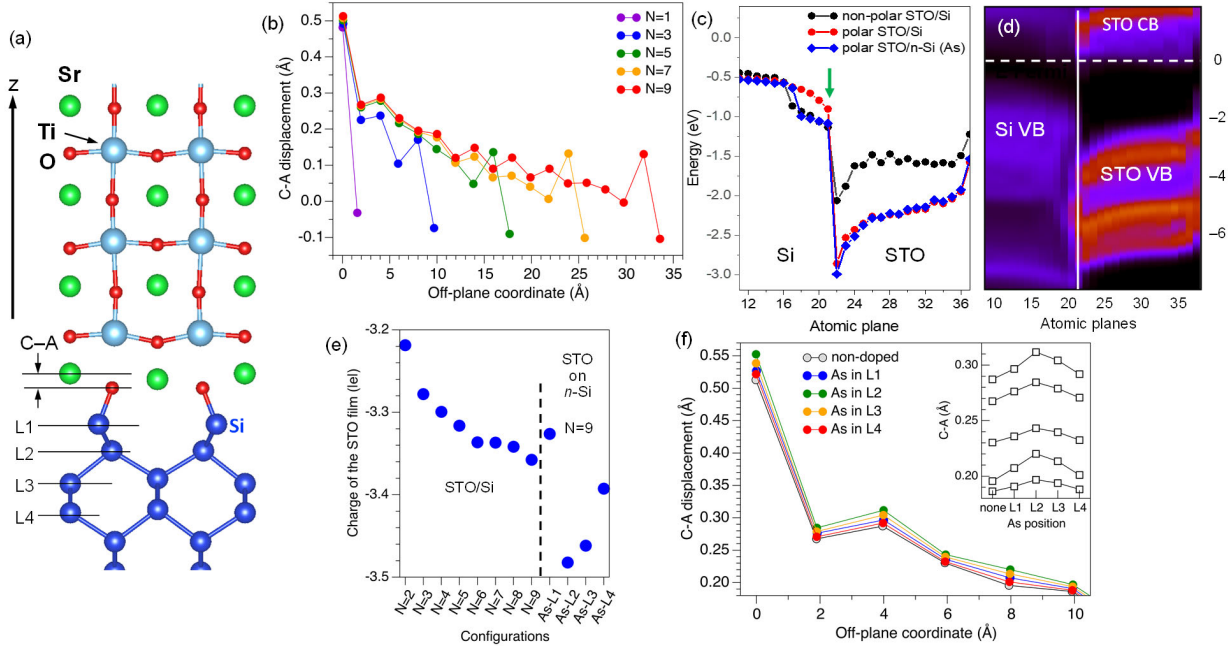

FIG. 5. (a) Structural model of the STO/Si interface; C-A shows the separation between the cation (C) and anion (A) atomic planes used as the measure of polarization in the STO film; L1 – L4 label Si atomic planes near the interface where the As impurity is located. (b) Polarization of the STO film as a function of film thickness $N$ (in u.c.) and distance from the interface. (c) Valence band edge profiles across the interface (arrow) for various STO/Si heterojunctions. The n-type As dopant is situated in row L2. (d) DOS projected on each atomic plane of the polar undoped STO/Si heterojunction. (e) Charge of the STO film calculated as a sum of atomic charges of the Sr, Ti, and O. (f) The effect of As doping on the polarization in the 9 u.c., thick STO film ($N = 9$). Inset: polarization in the planes 2–6 depending on the location of the As impurity.

We find that the polar heterojunction exhibits a larger valence band offset than the non-polar heterojunction, as shown in Fig. 5(c). Figure 5(d) shows the projected total DOS corresponding to the atomic planes of Si, SrO, and TiO$_2$ for the polar heterojunction that is 9 u.c. thick. Early studies identified the "interface dipole" to be the polar Si-O$^{2-}$ bonds [25]. In this regard, the Si-O$^{2-}$ bond length is shorter in the polar heterojunction than the non-polar. Heuristically, the Si-O$^{2-}$ bonds can be compared to a capacitor, in which shortening the Si-O$^{2-}$ bond length is analogous to reducing the separation between capacitor plates, thereby changing the potential. Later DFT studies, however, showed that the "interface dipole" arises from polarization in all the bonds across the interface, including those involving Sr [6, 8]. Within this expanded definition, the magnitude of the valence band offset was found to be directly correlated to the density of bonds in the interfacial region [8]. In agreement with these studies, we find that the presence of polarization not only shortens the Si$O^{2-}$ bond length, but also enhances bonding, as quantified by the number of electrons that STO gains from Si. For example, the total atomic charge over the 9 u.c. thick STO film in the polar heterojunction is -3.35 |e| per supercell, which far exceeds the -1.7 |e| per supercell in the 9 u.c. thick STO of the non-polar heterojunction. Finally, the polarization of STO also gives rise to a more pronounced, -z oriented electric field in the film that compensates or screens the +z oriented electric field across the interfacial Si-$O^{2-}$ region (Figs. 5(c) and 5(d)].

As n-type dopants are introduced into Si, our DFT models indicate that the total number of electrons in STO increases, consistent with the transfer of itinerant electrons observed in experiment. While it is experimentally well-established that O impurities can act as donors in Si, consensus as to how they do so has not been established amongst the many mechanisms that have been proposed [24, 26-29]. Thus, given the challenges of directly modeling O impurities as donors, we instead capture their effects by introducing an As$_{Si}$ defect. We introduced a substitutional As$_{Si}$ dopant with the concentration of $1.5 \times 10^{21}$ cm$^{-3}$ (one As atom per supercell) and varied its location with respect to the interface by placing it in Si atomic layers L1–L4 (Fig. 5(a)]. In the case of non-doped Si, the amount of charge transferred to STO increases with the thickness of STO, nearly saturating for a 9 u.c. film ($N = 9$ in Fig.5(e)]. Thus, we take the heterojunction with 9 u.c. STO as representative of the "bulk limit" of STO and introduce As$_{Si}$ dopants. If As is located in the outermost Si plane (L1), i.e., forms an interfacial As-O bond, an insignificant charge redistribution occurs between Si and STO. However, if As is fully coordinated by the Si atoms, the additional electron at the bottom of the conduction band of Si is partially transferred into the STO film. Such transfer is apparent for the As located in planes 2, 3, and 4 from the interface (L2, L3, and L4 in Fig. 5(a), respectively) and the magnitude of transferred charge decreases with increasing distance from the interface (Fig. 5(e)].

Using the simple heuristic definition of the interface dipole, we would expect that the space charge comprised of the transferred electrons in the STO along with the positive ionized donors in the Si would generate an electric field across the interface that "compresses" the Si-$O^{2-}$ dipole, thereby increasing the valence band offset. However, we find our DFT models predict that the valence band offset is largely invariant with As doping in Si, as shown in Fig. 5(c) for the As in the L2 configuration. Instead, our models predict an increase in the polarization of STO that largely mirrors the increase of electrons transferred from Si as a function of $As_{Si}$ location, as shown in Fig. 5(f). The enhanced polarization is strongest if the As impurity is located in the second Si plane from the interface (L2) and becomes less pronounced if As is in the 4$^{th}$ plane from the interface (L4).

To understand the presence (absence) of an enhanced valence band offset in experimental (theoretical) heterojunctions under charge transfer, we discuss differences between experiment and theory. Besides being much thicker, the 12 nm (~ 31 u.c.) thick STO films are structurally relaxed due to the formation of dislocations at steps on the Si(100) surface [2, 21]. In contrast, the 9 u.c. STO film modelled by DFT is coherently strained to the Si. Consequently, the change in polarization in the experimental heterojunction, shown in Fig. 1(e), stems from structural relaxation, whereas the change in polarization in the STO of the DFT model shown in Fig. 5(b) is driven by competition between oppositely oriented interface and surface polarizations. Due to this relaxation, the ions in the STO of experimental heterojunctions cannot "coherently" respond to the presence of transferred itinerant electrons through further polarization. Thus, we argue that the increase in valence band offset observed in experiment is a consequence of the inability of the lattice to compensate the effects of the transferred charge. We also note that DFT underestimates the size of bandgaps, which renders band offsets, electron transfer to the conduction band of STO, and the transition from type-II to type-III alignment difficult to capture (Fig. 5(d)). Nevertheless, our DFT models help elucidate the relationship between the interfacial dipole, polarization and band offset.

To provide further experimental evidence of the tunability of band alignment with space charge, we have also performed thickness dependent studies. Figures 6(a) and 6(b) show $R_s$ and $R_{xy}$ for heterojunctions with STO thicknesses of 4 nm and 8 nm. For reference, $R_s$ from the 12 nm thick STO/CZ-Si heterojunction is also shown. Note that the temperature at which a hole-gas forms in the Si progressively decreases as the thickness of the STO layer increases. ToF-SIMS shown in Fig. 6(c) reveals a systematic increase in O impurities in Si as the thickness of STO increases. Figures 6(d) and 6(e) show the Si 2p and Sr 3d core-level spectra and corresponding fits for the 4 nm and 8 nm thick STO/CZ-Si heterojunctions, respectively. These fits reveal a slight enhancement in the offset between the valence bands of STO and Si as the thickness of STO increases (Figs. 6(f) and 6(g)]. The thickness dependence of the band offsets explains the type-II deduced from all earlier studies of STO films grown on doped Si wafers, which were all performed on ultra-thin films that were < 2 nm thick [30, 31]. Also note that the Si in the 12 nm STO/FZ-Si heterojunction contains less O donors than the Si in the 4 nm STO/CZ-Si, indicating that the former is below the threshold density to induce a type-II to type-III transition (Fig. 6(c)]. Figure 7(a) summarizes the change in valence band offset as a function of integrated $^{18}O^-$ ToF-SIMS counts, which we define as the integrated area under the curve representing the $^{18}O^-$ signal of each heterojunction to a depth of 6 nm (Fig. 7(b)].

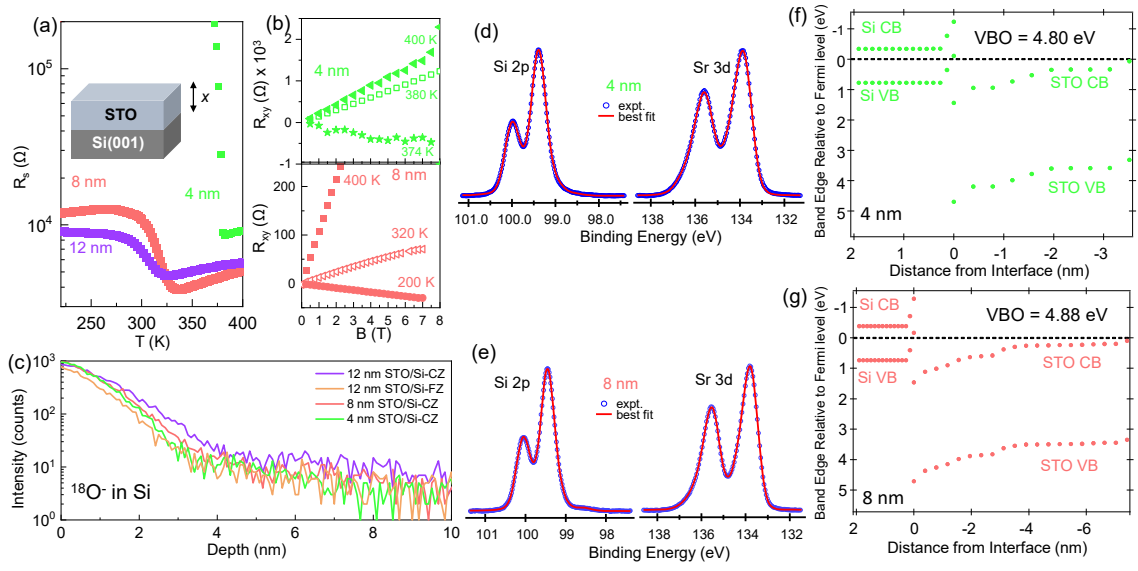

FIG. 6. (a) $R_s$ for the 4 nm, 8 nm, and 12 nm STO/CZ-Si heterojunctions. (b) $R_{xy}$ for the 4 nm and 8 nm thick STO/CZ-Si heterojunctions. (c) ToF-SIMS depth profiles of $^{18}O^-$ in Si for the 4 nm, 8 nm, 12 nm STO/Si-CZ heterojunctions. The data for the 12 nm STO/FZ-Si heterojunction is also shown for comparison. (d) and (e) Fits to the Si 2p and Sr 3d spectra for the 4 nm and 8 nm STO/CZ-Si heterojunctions, respectively. (f) and (g) The corresponding band-offsets for the 4 nm and 8 nm STO/CZ-Si heterojunctions, respectively.

The band offsets can also be tuned by introducing additional n-type donors in Si. Figures 8(c) and 8(d) show core-level spectra obtained on 12 nm thick STO films grown on n-type Si wafers that contain phosphorus donors that are uniformly distributed throughout the bulk, with concentrations of ~ $10^{15} - 10^{16}$ cm$^{-3}$, and $10^{17} - 10^{18}$ cm$^{-3}$, respectively. Note the decrease in intensity of the lower valence Ti features (Ti$^{3+}$, Ti$^{2+}$, etc.) with increasing density of n-type donors in Si, indicating that fewer electrons are transferred to STO. This decrease in transferred electrons is corroborated in Hall measurements shown in Fig. 8(b), as the STO grown on n-type Si ($10^{15} - 10^{16}$ cm$^{-3}$) exhibits a $n_{e(STO)}$ ~ $3.2 \times 10^{14}$ cm$^{-2}$, which is lower than the $n_{e(STO)}$ ~ $1 \times 10^{15}$ cm$^{-2}$ of the STO/CZ-Si heterojunction, in which both are compared at ~ 100 K. Here, we note that the bulk of the phosphorus doped n-type wafers shunts transport of the film and interface; hence only the $R_s$ (Fig. 8(a)) and $R_{xy}$ data below the temperature at which the bulk Si carriers freeze-out are analyzed. The STO grown on n-type Si with a higher dopant density ($10^{17} - 10^{18}$ cm$^{-3}$) is so resistive that $R_s$ and $R_{xy}$ could not be reliably measured below the temperature at which freeze-out of bulk Si carriers occurs. The presence of additional n-type donors dilutes the density of positive space charge near the interface, as itinerant electrons in the bulk of the Si screen the positive ionized donors associated with O, thereby effectively enhancing the screening length. As the density of positive space charge is diluted, the electric field across the interface is weakened, and the offset between the valence bands of STO and Si decreases, as shown in the band profiles of Figs. 8(e) and 8(f), for the STO/n-type CZ-Si heterojunctions of concentrations ~ $10^{15} - 10^{16}$ cm$^{-3}$ and $10^{17} - 10^{18}$ cm$^{-3}$, respectively.

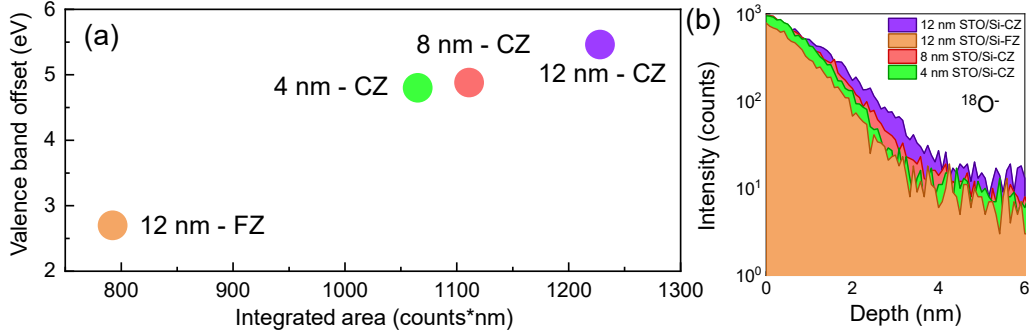

FIG. 7. (a) Valence band offset versus integrated area under $^{18}O^-$ profile, to a depth of 6 nm. (b) ToF-SIMS data showing integrated intensity with depth used to determine the integrated area.

## II. CONCLUSION

In summary, we demonstrate that the interfacial dipole at semiconductor-crystalline oxide interfaces can be tuned through space charge, thereby enabling band-alignment to be altered via doping. The ability to alter band alignment through doping sets hybrid heterojunctions apart from conventional heterojunctions. In the latter, changes in band offsets under doping are relatively modest, as interfacial dipoles are comparatively weak, as are electric fields due to space charge, given the longer screening lengths [11]. Tuning band-alignment with doping complements traditional approaches of altering band-alignment through composition [32], and opens new pathways to engineer functional behavior in hybrid heterojunctions.

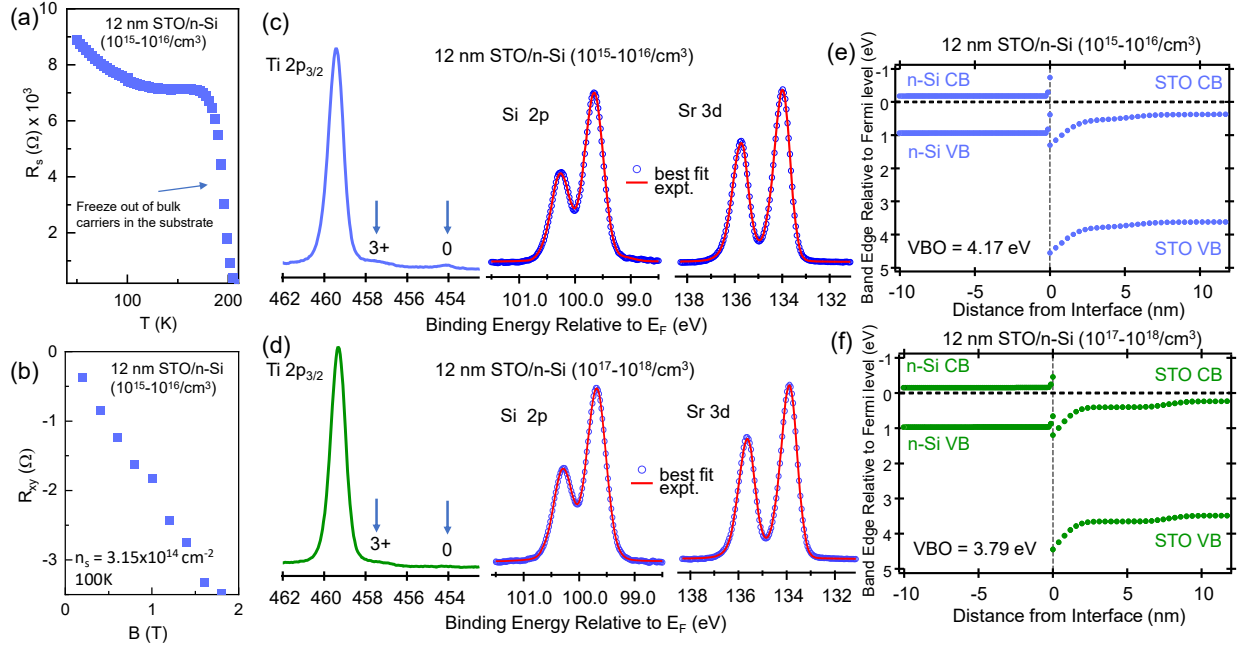

FIG. 8. (a) and (b) $R_s$ and $R_{xy}$ for the 12 nm STO/n-Si ($10^{15}$-$10^{16}$ cm$^{-3}$) heterojunction, respectively. (c) Ti 2p, Si 2p and Sr 3d core-level spectra for the 12 nm STO/n-Si ($10^{15}$-$10^{16}$ cm$^{-3}$) heterojunction. Fits to the Si 2p and Sr 3d spectra are also shown. (d) Ti 2p, Si 2p and Sr 3d core-level spectra for the 12 nm STO/n-Si ($10^{17}$-$10^{18}$ cm$^{-3}$) heterojunction. Fits to the Si 2p and Sr 3d spectra are also shown. (e) and (f) The band-offsets across the STO/n-Si ($10^{15}$-$10^{16}$ cm$^{-3}$) and STO/n-Si ($10^{17}$-$10^{18}$ cm$^{-3}$) heterojunctions, respectively.

## ACKNOWLEDGEMENTS

This work was supported by the National Science Foundation (DMR-1508530). S.A.C. and P.V.S. carried out the HAXPES data analysis and DFT modelling with support from the U.S. Department of Energy, Office of Science, Division of Materials Sciences and Engineering under Award #10122 to PNNL. PNNL is a multiprogram national laboratory operated for the U.S. Department of Energy (DOE) by Battelle Memorial Institute under Contract No. DE-AC05-76RL0-1830. STEM sample preparation was performed at the Environmental Molecular Sciences Laboratory (EMSL), a national scientific user facility sponsored by the DOE's

Biological and Environmental Research program and located at PNNL. Some STEM imaging was performed in the Radiological Microscopy Suite (RMS) located in the Radiochemical Processing Laboratory (RPL) at PNNL. SuperSTEM is the UK National Research Facility for Advanced Electron Microscopy, Funded by the Engineering and Physical Sciences Research Council. We thank Diamond Light Source for the access to Beamline I09 (via Beamtimes SI25582-2 and CM26487-1). JML and ANP acknowledge support from the National Science Foundation (DMR- 1350273). STEM imaging was performed in part at the Analytical Instrumentation Facility (AIF) at North Carolina State University, which is supported by the State of North Carolina and the National Science Foundation (award number ECCS-2025064). The AIF is a member of the North Carolina Research Triangle Nanotechnology Network (RTNN), a site in the National Nanotechnology Coordinated Infrastructure (NNCI). J.R.P., R.K.S., and A.X.G. acknowledge support from the U.S. Department of Energy, Office of Science, Office of Basic Energy Sciences, Materials Sciences, and Engineering Division under Award DE-SC0019297. The electrostatic photoelectron analyzer for the lab-based HAXPES measurements at Temple University was acquired through an Army Research Office DURIP grant (Grant W911NF-18-1-0251).